\documentclass[]{pasj01}
\draft

\begin{document} 
\Received{2015/09/28}
\Accepted{2016/01/17}

\title{Search for Soft X-ray Flashes at Fireball Phase
of Classical/Recurrent Novae using MAXI/GSC data}
				   
\author{Mikio \textsc{Morii}\altaffilmark{1}}
\altaffiltext{1}{Research Center for Statistical Machine Learning,
The Institute of Statistical Mathematics, \&
Core Research for Evolutionary Science and Technology
(CREST), Japan Science and Technology Agency (JST),
10-3 Midori-cho, Tachikawa, Tokyo 190-8562, Japan}
\email{morii@ism.ac.jp}
\author{Hitoshi \textsc{Yamaoka}\altaffilmark{2}}
\altaffiltext{2}{Department of Physics, Kyushu University,
6-10-1 Hakozaki, Fukuoka-shi Higashi-ku, Fukuoka 812-8581, Japan}
\author{Tatehiro \textsc{Mihara}\altaffilmark{3}}
\author{Masaru \textsc{Matsuoka}\altaffilmark{3}}
\altaffiltext{3}{MAXI Team,
Institute of Physical and Chemical Research (RIKEN),
2-1 Hirosawa, Wako, Saitama 351-0198, Japan}
\author{Nobuyuki \textsc{Kawai}\altaffilmark{4}}
\altaffiltext{4}{Department of Physics, Tokyo Institute of Technology,
Ookayama 2-12-1, Meguro-ku, Tokyo 152-8551, Japan}

\KeyWords{(stars:) novae, cataclysmic variables;
(stars:) white dwarfs; X-rays: bursts; methods: data analysis
} 

\maketitle

\begin{abstract}
We searched for precursive soft X-ray flashes (SXFs) associated
with optically-discovered classical or recurrent novae
in the data of five-years all-sky observations with
 Gas Slit Camera (GSC)
of the Monitor of All-sky X-ray Image (MAXI).
We first developed a tool to measure fluxes of point sources
by fitting the event distribution with the model
that incorporates the point-spread function (PSF-fit)
to minimize the potential contamination from nearby sources.
Then we applied the PSF-fit tool  to 40 classical/recurrent
novae that were discovered  in optical observations
from 2009 August to 2014 August.
We  found no precursive SXFs
with significance above $3 \sigma$ level
in the energy range of 2$-$4 keV
between $t_{d}-10$d and $t_{d}$,
where $t_{d}$ is the date when each nova was discovered. 
We obtained the upper limits for the bolometric luminosity of SXFs,
and compared  them with  the theoretical prediction
and that observed for MAXI J0158$-$744.
This result could constrain the population of massive
white dwarfs with a mass of roughly 1.40 solar mass, or larger,
in binary systems.
\end{abstract}

\section{Introduction}

Both classical and  recurrent novae are triggered by thermonuclear
runaways,  which last for $\sim$100~s at the surface of 
white dwarfs.  Subsequently, the optical flux increases  by
6 or more magnitudes, followed by an eventual decline to quiescence
\citep{Warner 1995}.
At the time of thermonuclear runaways,
an emission in the ultraviolet to soft X-ray bands  that lasts for only a few hours
is predicted  and is
called ``fireball phase'' \citep{Starrfield+2008, Krautter 2008}.
The fireball phase is  predicted to happen
a few days before the beginning of the optical nova phase,
although it has not been detected yet from any nova system.

In fact, MAXI/GSC did discover an extraordinarily luminous
soft X-ray flash (SXF) MAXI J0158$-$744 \citep{Li+2012, Morii+2013}.
The MAXI J0158$-$744 system was considered to be in the fireball phase,
although there was no association with a usual classical/recurrent nova.
This SXF is characterized  with a soft X-ray spectrum,
a short duration ($\Delta T_d$; $1.3 \times 10^3$~s
$< \Delta T_d < $ $1.10 \times 10^4$~s),
a rapid rise ($< 5.5 \times 10^3$~s), and
a huge peak luminosity of $2 \times 10^{40}$ erg s$^{-1}$
in the $0.7-7.0$ keV band.
Although the characteristics of this flash are
 very different from those of usual novae,
the soft X-ray emission was successfully interpreted
as the fireball phase of a new kind of novae
on a very massive white dwarf
\citep{Morii+2013}.
 Indeed, this hypothesis was further supported  with a detailed simulation 
performed by \citet{Ohtani+2014}.
The soft X-ray emission  during the fireball phase is
a photospheric emission,  and accordingly, the spectral shape is
basically a blackbody.
\citet{Morii+2013} also concluded that 
the small increase in the flux observed in the optical
counterpart of MAXI J0158$-$744 did not originate
from a photospheric emission as usual novae,
but is due to the emission from the disk around
the Be companion star, which is a reprocess,
originated from the photospheric soft X-ray emission.
 Now, unlike MAXI J0158$-$744, the companion star in the white-dwarf binaries that generate novae is in general not a Be star.
 Then, it is unlikely any detectable increase in the optical flux would occur
 during SXF of a nova, where the companion star is not a Be star.

 It has been argued in many theoretical studies that
white dwarfs can acquire a mass close to
or over the Chandrasekhar limit via differential rotation
(e.g. \citet{Yoon Langer 2004} and \citet{Hachisu+2012})
or strong internal magnetic fields
(e.g. \citet{Das Mukhopadhyay 2012} and \citet{Franzon Schramm 2015}).
However, the theoretical works have not caught up with the observational
discovery of the new type of nova from a very massive white dwarf
(MAXI J0158$-$477; \citet{Morii+2013}).
At the same time, we need more observational samples
to study and understand the phenomenon
and its background science.
Here we present the result of the systematic search for 
nova explosions from very massive white dwarfs.

In this paper, we perform the systematic search for
the fireball phase emission in the soft X-ray  band
for usual classical/recurrent novae,
using the association of optically discovered novae 
as predicted by \citet{Starrfield+2008} and \citet{Krautter 2008}.
We should note that the SXF without optical-nova phase like
MAXI J0158$-$744 are out of our samples and hence would not be detected in this search.
However, intermediate objects that possess both
the fireball phase with a SXF emission 
and a optical nova phase are expected to be found,
if such objects are actually present.

\section{Observation and Analysis}
\label{sec: obs ana}

MAXI (Monitor of All-sky X-ray Image; \cite{Matsuoka+2009})
is an all-sky X-ray monitor,
which is operated on the Japanese Experiment Module (KIBO)
on the International Space Station (ISS).
MAXI carries Gas Slit Camera (GSC; \cite{Mihara+2011, Sugizaki+2011}),
which scans the almost  entire sky every $\sim 92$ minutes
through the long and narrow fields of view (FoVs) of
\timeform{1.5D}$\times$\timeform{160D}.
The scan duration for a point source is $40-150$~s
\citep{Sugizaki+2011}.
GSC with its gas proportional counters is sensitive for $2 - 30$ keV. 
From the start of the operation on 2009 August 15 up to the present day,
GSC has almost continuously monitored the whole sky.
All  the data have been stored in
Japan Aerospace Exploration Agency (JAXA)
and 
Institute of Physical and Chemical Research (RIKEN).

Table \ref{table:nova1} shows the classical/recurrent novae
discovered in the period from 2009 August 15 to  2014 August 15, taken
 from the reports of
the International Astronomical Union Circulars (IAUC)  and
Central Bureau Electronic Telegrams (CBET).
MAXI/GSC observed these fields of the sky at  the period
 of the discovery in the optical wavelength.
 Accordingly, MAXI/GSC can search the precursive activity
in the X-ray band for these novae.

We started the data analysis from the event data
of MAXI/GSC stored in every day and every camera in the \texttt{fits} format.
For every source, we extracted events from a circular sky region
centered on the target with a radius of \timeform{8D},
 using \texttt{mxextract}.
We also calculated the time variation of the effective area
in every 1~s for the point source
determined by slat-slit collimator of MAXI/GSC,
 using \texttt{mxscancur}.
We made good time intervals (GTIs) to include this duration
and to remove the duration when the counter was off.
We also removed the scans  during which the solar paddle of the ISS 
 obscured the FoV of the GSC counter.
The selected event data in \texttt{fits} format were
converted to \texttt{root}
\footnote{https://root.cern.ch/drupal/} format
 to enhance visualization of the data.
For every scan, we measured the flux of the target
in  the unit of counts s$^{-1}$ cm$^{-2}$ in the $2-4$ keV band,
 using  the tool  named ``PSF-fit''.
The details of  the ``PSF-fit'' are described in Appendix \ref{sec:appendix}.
The spectrum file and the corresponding response file were
made for every scan by using \texttt{xselect} and \texttt{mxgrmfgen},
respectively.

We made the light curves of the sources listed in
Table \ref{table:nova1} in every GSC scan,
 using the PSF-fit tool from $t_{d} -50$~d to $t_{d} +50$~d,
where $t_{d}$ is
the  time, when each nova was discovered in the optical wavelength.
We also calculated the upper limits for the fluxes
in 90\% confidence level (C.L.) for the scans with non-detection,
using the same tool.

\begin{longtable}{lrrlcl}
\caption{Forty-four classical/recurrent novae discovered
in the optical wavelengths  during five years
from 2009 August to 2014 August}
\label{table:nova1}
\hline\hline
\multicolumn{1}{c}{Nova}      &
\multicolumn{1}{c}{R.A.$^1$}  &
\multicolumn{1}{c}{Dec.$^2$}  &
\multicolumn{1}{c}{Discovery} &
\multicolumn{1}{c}{Class$^3$} &
\multicolumn{1}{l}{Reference$^4$}   \\
          &
\multicolumn{1}{c}{deg}           &
\multicolumn{1}{c}{deg}           &
\multicolumn{1}{c}{Date (UT)$^5$} &
                                  & 
                                  \\
\hline
\endfirsthead
V2672 Oph       & $264.582$ &  $-26.737$  &    2009-08-16.515       &    CN     &    IAUC 9064                 \\
V5584 Sgr       & $277.887$ &  $-16.319$  &    2009-10-26.439       &    CN     &    IAUC 9089                 \\
V496 Sct        &  $280.940$ &  $-7.612$  &    2009-11-08.370       &    CN     &    IAUC 9093, CBET 2008      \\
KT Eri          &  $71.976$ &  $-10.179$  &    2009-11-25.545       &    CN     &    IAUC 9098                 \\
V1722 Aql       &  $288.541$ &  $+15.276$  &    2009-12-14.40        &    CN     &    IAUC 9100                 \\
V2673 Oph       &  $264.921$ &  $-21.663$  &    2010-01-15.857       &    CN     &    IAUC 9111                 \\
V5585 Sgr       &  $271.862$ &  $-29.012$  &    2010-01-20.72        &    CN     &    IAUC 9112                 \\
U Sco           &  $245.628$ &  $-17.879$  &    2010-01-28.4385      &    RN     &    IAUC 9111                 \\
V2674 Oph       &  $261.634$ &  $-28.827$  &    2010-02-18.845       &    CN     &    IAUC 9119                 \\
V1310 Sco       &  $256.531$ &  $-37.241$  &    2010-02-20.857       &    CN     &    IAUC 9120                 \\
V407 Cyg        &  $315.541$ &  $+45.776$  &    2010-03-10.797       &    SyN    &    CBET 2199                 \\
V5586 Sgr       &  $268.262$ &  $-28.205$  &    2010-04-23.782       &    CN     &    IAUC 9140                 \\
V1311 Sco       &  $253.805$ &  $-38.063$  &    2010-04-25.788       &    CN     &    IAUC 9142                 \\
V1723 Aql       &  $281.910$ &  $-3.787$  &    2010-09-11.485       &    CN     &    IAUC 9167                 \\
V5587 Sgr       &  $266.943$ &  $-23.587$  &    2011-01-25.86        &    CN     &    IAUC 9196                 \\
V5588 Sgr       &  $272.589$ &  $-23.092$  &    2011-03-27.832       &    CN     &    IAUC 9203, CBET 2679      \\
T Pyx           &  $136.173$ &  $-32.380$  &    2011-04-14.2931      &    RN     &    IAUC 9205, CBET 2700      \\
V1312 Sco       &  $253.789$ &  $-38.635$  &    2011-06-01.40        &    CN     &    IAUC 9216, CBET 2735      \\
PR Lup          &  $223.583$ &  $-55.084$  &    2011-08-04.73        &    CN     &    IAUC 9228, CBET 2796      \\
V1313 Sco       &  $249.179$ &  $-41.546$  &    2011-09-06.37        &    CN     &    IAUC 9233, CBET 2813      \\
V965 Per        &  $47.818$ &  $+37.084$  &    2011-11-07.75        &    CN     &    IAUC 9247                 \\
V834 Car        &  $162.582$ &  $-64.113$  &    2012-02-26.543       &    CN     &    IAUC 9251, CBET 3040      \\
V1368 Cen       &  $205.289$ &  $-58.255$  &    2012-03-23.386       &    CN     &    IAUC 9260, CBET 3073      \\
V2676 Oph       &  $261.529$ &  $-25.862$  &    2012-03-25.789       &    CN     &    IAUC 9259, CBET 3072      \\
V5589 Sgr       &  $266.367$ &  $-23.090$  &    2012-04-21.01123     &    CN     &    IAUC 9259, CBET 3089      \\
V5590 Sgr       &  $272.766$ &  $-27.291$  &    2012-04-23.689       &    CN     &    IAUC 9259, CBET 3140      \\
V2677 Oph       &  $264.983$ &  $-24.795$  &    2012-05-19.484       &    CN     &    IAUC 9260, CBET 3124      \\
V1324 Sco       &  $267.725$ &  $-32.622$  &    2012-05-22.80        &    CN     &    CBET 3136                 \\
V5591 Sgr       &  $268.107$ &  $-21.439$  &    2012-06-26.5494      &    CN     &    IAUC 9259, CBET 3156      \\
V5592 Sgr       &  $275.114$ &  $-27.741$  &    2012-07-07.4986      &    CN     &    IAUC 9259, CBET 3166      \\
V5593 Sgr       &  $274.904$ &  $-19.128$  &    2012-07-16.512       &    CN     &    IAUC 9259, CBET 3182      \\
V959 Mon        &  $99.911$ &  $+5.898$  &    2012-08-07.8048      &    CN     &    IAUC 9259, CBET 3202      \\
V1724 Aql       &  $283.146$ &  $+0.312$  &    2012-10-20.4294      &    CN     &    IAUC 9259, CBET 3273      \\
V809 Cep        &  $347.020$ &  $+60.781$  &    2013-02-02.4119      &    CN     &    IAUC 9260, CBET 3397      \\
V1533 Sco       &  $263.498$ &  $-36.106$  &    2013-06-03.6146      &    CN     &    IAUC 9260, CBET 3542      \\
V339 Del        &  $305.878$ &  $+20.768$  &    2013-08-14.5843      &    CN     &    IAUC 9258, CBET 3628      \\
V1830 Aql       &  $285.639$ &  $+3.255$  &    2013-10-28.4571      &    CN     &    IAUC 9263, CBET 3691, CBET 3708   \\
V556 Ser        &  $272.264$ &  $-11.210$  &    2013-11-24.3835      &    CN     &    IAUC 9264, CBET 3724      \\
V1369 Cen       &  $208.696$ &  $-59.152$  &    2013-12-02.692       &    CN     &    IAUC 9265, CBET 3732      \\
V5666 Sgr       &  $276.286$ &  $-22.601$  &    2014-01-26.857       &    CN     &    IAUC 9269, CBET 3802      \\
V745 Sco        &  $268.843$ &  $-33.250$  &    2014-02-06.694       &    RN     &    CBET 3803                 \\
V962 Cep        &  $313.599$ &  $+60.285$  &    2014-03-08.7917      &    CN     &    IAUC 9270, CBET 3825      \\
V1534 Sco       &  $258.945$ &  $-31.475$  &    2014-03-26.8487      &    CN     &    IAUC 9273, CBET 3841      \\
V2659 Cyg       &  $305.426$ &  $+31.058$  &    2014-03-31.7899      &    CN     &    IAUC 9271, CBET 3842      \\
\hline
\multicolumn{6}{l}{$^1$Right ascension. $^2$Declination. } \\
\multicolumn{6}{l}{$^3$CN: Classical Nova; RN: Recurrent Nova; SyN: Symbiotic Nova} \\
\multicolumn{6}{l}{$^4$IAUC: The International Astronomical Union Circulars,} \\
\multicolumn{6}{l}{CBET: Central Bureau Electronic Telegrams} \\
\multicolumn{6}{l}{$^5$(Year)-(Month)-(Day)}
\end{longtable}


\begin{longtable}{llccll}
\caption{Results of our search for soft X-ray flashes for
the forty novae in Table \ref{table:nova1}}
\label{table:nova2}
\hline\hline
\multicolumn{1}{c}{Nova$^1$}      &
\multicolumn{1}{c}{$N_{\rm H}$$^2$ } &
\multicolumn{1}{c}{Distance} &
\multicolumn{1}{c}{U.L.$^3$} &
\multicolumn{1}{l}{$N_{\rm s}$$^4$}    &
\multicolumn{1}{l}{Reference for distance} \\
          &
\multicolumn{1}{c}{cm$^{-2}$}      &
\multicolumn{1}{c}{kpc}            &
\multicolumn{1}{c}{ave (min -- max)} &
                                   &
                                   \\
\hline
\endfirsthead
V2672 Oph       &  $4.18 \times 10^{21}$ &  $19  \pm 2$    & 2.51 (0.71 -- 4.24)  & 22    &  \citet{Takei+2014} \\
V5584 Sgr       &  $3.96 \times 10^{21}$ &  $6.3 \pm 0.5$  & 1.20 (0.38 -- 3.28)  & 134   &  \citet{Raj+2015}   \\
V496 Sct        &  $6.81 \times 10^{21}$ &  $2.9 \pm 0.3$  & 2.22 (0.74 -- 4.75)  & 45    &  \citet{Raj+2012}   \\
KT Eri          &  $5.52 \times 10^{20}$ &  $6.6 \pm 0.8$  & 1.71 (0.61 -- 6.03)  & 109   &  \citet{Imamura Tanabe 2012} \\
V1722 Aql       &  $9.14 \times 10^{21}$ &  $5   $         & 1.97 (0.63 -- 6.01)  & 112   &  \citet{Munari+2010} \\
V2673 Oph       &  $2.84 \times 10^{21}$ &  $7.4 $         & 1.42 (0.43 -- 4.38)  & 77    &  \citet{Munari Dallaporta 2010} \\
V5585 Sgr       &  $2.75 \times 10^{21}$ &  N/A            & 1.76 (0.50 -- 4.36)  & 44    &  \\
V2674 Oph       &  $4.12 \times 10^{21}$ &  $ 9  $         & 1.75 (0.67 -- 3.96)  & 30    &  \citet{Munari Dallaporta Ochner 2010} \\
V407 Cyg        &  $8.19 \times 10^{21}$ &  $2.7$          & 1.83 (0.56 -- 4.46)  & 52    &  \citet{Munari Margoni Stagni 1990} \\
V5586 Sgr       &  $9.60 \times 10^{21}$ &  N/A            & 2.79 (1.21 -- 5.64)  & 23    &  \\
V1311 Sco       &  $4.78 \times 10^{21}$ &  N/A            & 1.98 (0.52 -- 4.44)  & 20    &  \\
V1723 Aql       &  $1.23 \times 10^{22}$ &  $   6$         & 1.76 (0.49 -- 4.97)  & 140   &  \citet{Weston+2015} \\
V5587 Sgr       &  $4.09 \times 10^{21}$ &  N/A            & 1.56 (0.30 -- 3.67)  & 126   &  \\
V5588 Sgr       &  $5.47 \times 10^{21}$ &  $7.6$          & 2.15 (0.65 -- 5.67)  & 122   &  \citet{Munari+2015}  \\
T Pyx           &  $1.88 \times 10^{21}$ &  $3.5 \pm  1$   & 1.87 (0.43 -- 8.19)  & 35    &  \citet{Schaefer 2010} \\
V1312 Sco       &  $5.14 \times 10^{21}$ &  N/A            & 1.61 (0.47 -- 4.64)  & 118   &  \\
PR Lup          &  $4.82 \times 10^{21}$ &  N/A            & 1.25 (0.39 -- 3.76)  & 144   &  \\
V1313 Sco       &  $4.77 \times 10^{21}$ &  N/A            & 1.09 (0.26 -- 3.71)  & 55    &  \\
V965 Per        &  $1.20 \times 10^{21}$ &  N/A            & 3.56 (1.37 -- 8.18)  & 17    &  \\
V834 Car        &  $4.98 \times 10^{21}$ &  N/A            & 1.49 (0.38 -- 5.09)  & 121   &  \\
V1368 Cen       &  $4.30 \times 10^{21}$ &  N/A            & 1.37 (0.46 -- 3.50)  & 47    &  \\
V2676 Oph       &  $3.02 \times 10^{21}$ &  N/A            & 1.34 (0.36 -- 3.56)  & 89    &  \\
V5589 Sgr       &  $3.35 \times 10^{21}$ &  N/A            & 2.21 (0.47 -- 5.85)  & 111   &  \\
V5590 Sgr       &  $2.64 \times 10^{21}$ &  N/A            & 2.19 (0.70 -- 5.50)  & 107   &  \\
V2677 Oph       &  $3.38 \times 10^{21}$ &  N/A            & 3.46 (0.79 -- 11.27) &  44   &  \\
V1324 Sco       &  $4.35 \times 10^{21}$ &  $4.5$          & 2.56 (0.79 -- 6.05)  & 47    & \citet{Ackermann+2014} \\
V5591 Sgr       &  $4.27 \times 10^{21}$ &  N/A            & 2.31 (0.61 -- 6.59)  & 134   &  \\
V5592 Sgr       &  $1.49 \times 10^{21}$ &  N/A            & 2.01 (0.72 -- 5.96)  & 93    &  \\
V5593 Sgr       &  $6.54 \times 10^{21}$ &  N/A            & 2.31 (0.66 -- 5.97)  & 115   &  \\
V959 Mon        &  $6.29 \times 10^{21}$ &  $3.6$          & 2.68 (0.71 -- 9.77)  & 118   & \citet{Shore+2013} \\
V1724 Aql       &  $1.47 \times 10^{22}$ &  N/A            & 2.49 (0.61 -- 7.11)  & 133   &  \\
V809 Cep        &  $8.22 \times 10^{21}$ &  $6.5$          & 1.98 (0.65 -- 5.25)  & 93    & \citet{Munari+2014} \\
V1533 Sco       &  $7.42 \times 10^{21}$ &  N/A            & 2.10 (0.66 -- 4.36)  & 84    &  \\
V339 Del        &  $1.36 \times 10^{21}$ &  $4.2$          & 2.04 (0.64 -- 6.08)  & 140   & \citet{Shore 2013} \\
V1830 Aql       &  $1.23 \times 10^{22}$ &  N/A            & 3.08 (0.86 -- 8.20)  & 128   &  \\
V556 Ser        &  $4.87 \times 10^{21}$ &  N/A            & 2.54 (0.93 -- 6.70)  & 89    &  \\
V1369 Cen       &  $5.84 \times 10^{21}$ &  $2.4$          & 2.58 (0.74 -- 8.95)  & 83    & \citet{Shore+2014} \\
V962 Cep        &  $3.30 \times 10^{21}$ &  N/A            & 2.72 (1.11 -- 11.86) & 125  &  \\
V1534 Sco       &  $3.88 \times 10^{21}$ &  $13$           & 1.77 (0.79 -- 4.24)  & 22    & \citet{Joshi+2015} \\
V2659 Cyg       &  $4.92 \times 10^{21}$ &  N/A            & 2.27 (0.73 -- 6.13)  & 124   &  \\
\hline
\multicolumn{6}{l}{$^1$U Sco, V1310 Sco, V5666 Sgr and V745 Sco
listed in Table \ref{table:nova1} are removed (see text).} \\
\multicolumn{6}{l}{$^2$Total Galactic HI column density toward the source as  the average} \\
\multicolumn{6}{l}{of the values obtained by Leiden/Argentine/Bonn (LAB) map\citep{Kalberla+2005}} \\
\multicolumn{6}{l}{and DL map \citep{Dickey Lockman 1990}, calculated using HEASARC Web site:} \\
\multicolumn{6}{l}{http://heasarc.gsfc.nasa.gov/cgi-bin/Tools/w3nh/w3nh.pl} \\
\multicolumn{6}{l}{$^3$90\%C.L. upper limit for the flux in the 2$-$4 keV band for each scan
  in  the unit of $10^{-2}$ counts s$^{-1}$ cm$^{-2}$.} \\
\multicolumn{6}{l}{Average (minimum and maximum) fluxes among $N_{\rm s}$ scans are shown.} \\
\multicolumn{6}{l}{$^4$Number of GSC scans in the searched period.}
\end{longtable}

\section{Results}

For every nova listed in Table \ref{table:nova2},
no significant precursive SXF was
found  at more than $3 \sigma$ level
during the period between $t_{d}-10$d and $t_{d}$.
 Note that in this table, U Sco, V1310 Sco, V5666 Sgr and V745 Sco
listed in Table \ref{table:nova1} are removed due to the following reasons.
U Sco  is removed due to the occasional contamination of
the tail of the PSF of Sco X-1.
Although the angular distance between them
is not so small (2.4 deg), Sco X-1 is 
the brightest X-ray source in the  entire sky ($\sim 10$ Crab),
then the contamination becomes sometimes severe.
V1310 Sco  is also removed due to the contamination of
the nearby bright source GX 349+2, 
which is $0.8$ deg apart from V1310 Sco.
For V5666 Sgr and V745 Sco, there  was no GTI of GSC scans during
the searched periods.

\begin{figure}
 \begin{center}
   \includegraphics[width=6cm, angle=-90]{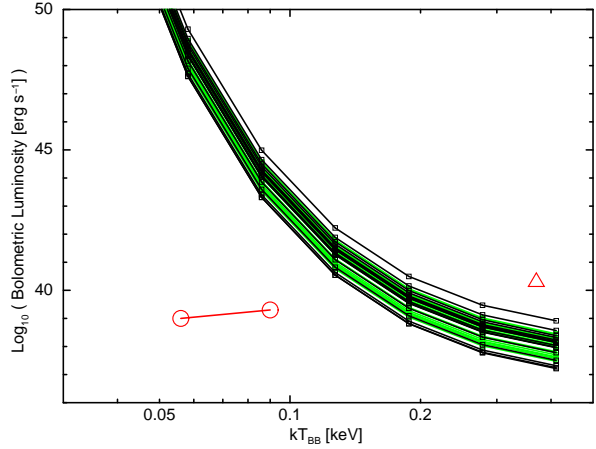} 	
 \end{center}
\caption{The 90\% C.L. upper limits for bolometric luminosity
of the fireball phase of classical/recurrent novae searched
for in this work (see Table \ref{table:nova2}).
 In the conversion
from the observed GSC flux in the $2-4$ keV band (counts s$^{-1}$ cm$^{-2}$)
to the unabsorbed bolometric luminosity (erg s$^{-1}$),
a blackbody spectrum is assumed,
 and the interstellar absorption $N_H$ and
 distance listed in Table \ref{table:nova2} are used.
For the sources with unknown distance, we assumed 5 kpc
as a typical distance for Galactic sources.
The upper limits for the sources
with known and unknown distances are
shown in black and green lines, respectively.
Horizontal and vertical axes are
the temperature of blackbody assumed in the flux conversion
in  the units of keV and
logarithm of bolometric luminosity, respectively.
The theoretical prediction of the fireball phase 
\citep{Starrfield+2008}  is  given
in red circle points connected with a solid line,
to be compared with the derived upper limits,
 where the left and right circle-points correspond to
the emission  in the fireball phase of white dwarfs
with respective 1.25 and 1.35 solar masses. 
The red triangle point is the observed value
for the fireball phase on MAXI J0158$-$744 \citep{Morii+2013}.
}\label{fig:lumkt}
\end{figure}

We then calculated the 90\%C.L. upper limit for the fluxes in every scan 
for the sources in  the unit of counts s$^{-1}$ cm$^{-2}$ in the $2-4$ keV band
(Table \ref{table:nova2}).
Since the expected times of the SXFs are unknown,
 the upper limits given in the table
 are the averaged values for all the scans between 
$t_{d}-10$d and $t_{d}$,  as well as the minimum and maximum.
From these averaged flux upper-limits, 
we calculated the upper limits for bolometric luminosity
of the sources, using the interstellar absorption 
$N_H$ and the distance (Table \ref{table:nova2}),
and assuming the temperature of blackbody spectrum for 
the fireball-phase emission.
 Figure~\ref{fig:lumkt} shows the derived upper limits.

\section{Discussion and Conclusion}

If SXFs are associated precursively with optical novae,
if the spectrum and flux are similar to
those of MAXI J0158$-$744, and 
if the nova  is located at the typical distance of 5 kpc,
then 
MAXI/GSC should detect these flashes in high significance (see Figure \ref{fig:lumkt}),
as long as the FoV of GSC  covers the direction during the SXF.
MAXI/GSC usually scans a specific direction
every 92 minutes.  Hence, the probability
that MAXI/GSC detects a SXF depends on
the duration of the SXF.

\begin{figure}
 \begin{center}
   \includegraphics[width=6cm, angle=-90]{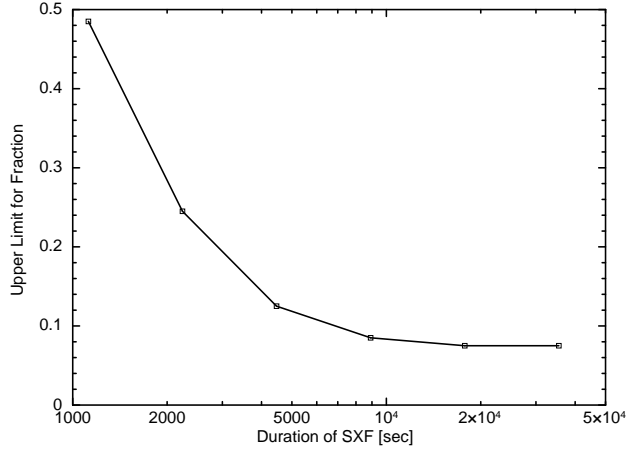} 	
 \end{center}
\caption{The 90\% C.L. upper limit for the fraction of novae
associated with a  SXF at the fireball phase ($R$)
as a function of the given
 duration of SXFs in  the unit of second (see text).}\label{fig:sec_width_ul}
\end{figure}

In our study, we performed $n = 40$ trials
of MAXI/GSC observations for SXFs and  found no detection. 
Statistically speaking, our experiment
is a sequence of independent Bernoulli trials
with a success probability of $R p_c^{(i)}$, where
 $R$ is the fraction of novae associated with SXF
at the fireball phase and 
$p_c^{(i)}$ is the probability
for MAXI/GSC to scan the direction of the $i$-th nova
during the search period between $t_d - 10$~d and $t_d$.
By denoting the detection or non-detection for the $i$-th trial
as $d_i = 1$ or $0$, respectively,
the joint probability of ${\bf d}$ throughout this experiment
is  given by
\begin{equation}
P({\bf d}; R, {\bf p_c}) = 
\prod_{i = 1}^n (R p_c^{(i)})^{d_i} (1 - R p_c^{(i)})^{1 - d_i},
\label{eq: joint prob}
\end{equation}
where ${\bf d} = (d_1, \cdots, d_n)$ and
${\bf p_c} = (p_c^{(1)}, \cdots, p_c^{(n)})$ are
vectors of the detection and the probability, respectively.
 Following our result of non-detection (${\bf d} = {\bf 0}$),
we calculated the upper limit for the fraction
$R$ to be in 90\% confidence level,
varying the duration of the SXF.
Figure \ref{fig:sec_width_ul} shows this upper limit
as a function of the duration of the SXF, where
 the limit was calculated  with the standard
method of interval estimation of a parameter by
inverting a likelihood-ratio test-statistic
(Section 9.2 in \cite{Casella Berger 2002}).
 Note that we evaluated the probability distribution
 of the function  in Equation \ref{eq: joint prob}
with Monte Carlo simulation in the calculation.

In Figure \ref{fig:lumkt}, 
the fireball phase observed on MAXI J0158$-$744
 is located at the extension of theoretical prediction
\citep{Starrfield+2008}.
It supports that the SXF observed on MAXI J0158$-$744
is the emission from the fireball phase.
The mass of the white dwarf of this source is estimated to
be at least  1.35 solar mass.
Furthermore, by naively drawing a straight line connecting the two  points 
of 1.35 solar mass (the right red circle) and
MAXI J0158$-$744 (red triangle), the crossing point 
between this line and  the average of the lines of the upper limits
implies a lower limit for the mass
of white dwarfs for which the SXFs are detectable
 with MAXI/GSC.
Assuming that the mass of MAXI J0158$-$744
is at the Chandrasekhar limit of white dwarfs
(1.44 solar mass), the mass of a white dwarf
 that is located at the crossing point is estimated
to be about 1.40 solar mass.
Therefore, the upper limit for the fraction
in Figure \ref{fig:sec_width_ul} can be
alternatively interpreted 
as the upper limit for the population of white dwarfs
in binary systems whose masses
are  about $1.40$ solar mass or larger.
The population of very massive white dwarfs
is important for estimating the rate of type Ia supernovae.
Our result could give a constraint for this population.

Future observations  with MAXI/GSC will detect more SXFs
 from novae in the fireball phases. 
In addition,  wide-field MAXI mission under planning 
\citep{Kawai+2014}
would improve the detection efficiency of SXFs
and would increase the sample of SXFs and very massive white dwarfs.
 In the light of these prospects, the theoretical study  of the fireball-phase emission
of novae on very massive white dwarfs,
especially from 1.35 solar mass to near the Chandrasekhar limit,
is encouraged.

\begin{ack}
We  thank the member of the MAXI operation team.
We also thank J. Shimanoe for the first search for transients
from optical novae. 
K. Sugimori, T. Toizumi and T. Yoshii
 kindly offered us a good help in developing the PSF-fit.
We are grateful for S. Ikeda on ISM, who advised for
the statistical treatment of the data analysis.
This work was partially supported by the Ministry of Education,
Culture, Sports, Science and Technology (MEXT),
Grant-in-Aid for Science Research 24340041.
\end{ack}

\appendix

\section{Flux measurement
by using the point-spread function --- PSF-fit}\label{sec:appendix}

X-ray fluxes of point sources are usually measured
 with aperture photometry,  and the standard processing for the MAXI data too follows it.  
The MAXI light-curves obtained  with  this method are  published in the archive 
 on the website of RIKEN
\footnote{http://maxi.riken.jp/top/}.
 In the standard MAXI data-processing, the source and background regions 
to extract events are
selected usually as simple circular and annulus regions
in the sky coordinates, respectively.
When the point-spread functions (PSFs)
of nearby sources interfere these regions,
the circular regions centered on these nearby sources are excluded.
If a target source  is located in a severely crowded region like the Galactic plane,
 measurement  of the flux based on such a simple selection of the regions is inadequate,
 because the PSF of a target  overlaps
with those of nearby sources and because the area of the background region
free from contamination of the nearby sources becomes too small to obtain the adequate statistics.

In this work, we developed a tool to measure
the fluxes for the point sources
by fitting event distribution around a target
with a model  that incorporates of
 the PSFs of the target and nearby sources.
 Both X-ray and non-X-ray backgrounds  are also  taken into account in this model.
 We  name this tool the PSF-fit.
 The following is the detailed description of the PSF-fit.\newline
 {\indent}We use event distribution in the detector coordinate
 rather than the sky coordinate,
 because the shape of a PSF is best described in
 the detector coordinate.
 In this coordinate, the shape of a PSF
 is  given by a triangular-shaped function
 in the time (scan) direction,
  whereas that in the anode wire direction
 is  given by a Gaussian shape \citep{Sugizaki+2011}.
 The triangular shape is well calibrated 
  \citep{Morii-Sugimori-Kawai 2011}.
 The width of the Gaussian shape (sigma) depends on counters,
 the position of the anode direction, and the high voltage
 (HV; 1650V or 1550V).
 We calibrated this dependence by measuring
 the width of the Gaussian shape for the bright point sources,
 Crab nebula, GRS 1915+105, and Cyg X-1,
 for the energy bands of $2-4$ and $4-10$ keV.
  Thus, we obtained the well-calibrated PSF model.

 We modeled the event distribution
 in  the detector coordinate ($t$, $b$) of a camera
 ($k = 0, \cdots ,11$; camera IDs) as follows:
 \begin{equation}
   F_{\rm model}(t, b; k)
   = \sum_{m \in S_k} f^{(m)}
   g_{\rm psf}^{(m)} (t, b; \alpha_m, \delta_m, k) + B(t, b; k),
 \end{equation}
 where $t$ is time (s),  $b$ is the position of anode, bex (mm), and
 ($\alpha_m$, $\delta_m$) are the sky coordinates 
 (Right ascension, Declination) of the target
 and nearby sources. $S_k$ is a set of sources,
 including the target and nearby sources ($m$: source IDs),
 and they are different  among cameras ($k$),
because the FoVs and condition of each camera are different.
 $f^{(m)}$ and $g_{\rm psf}^{(m)} (t, b)$ are the
source flux in  the unit of counts s$^{-1}$ cm$^{-2}$ and
the shape of the PSF in  the unit of cm$^{2}$ mm$^{-1}$.
$B(t, b; k)$ is  the shape of a background rate
in  the unit of counts s$^{-1}$ mm$^{-1}$,
for which we used a constant function in this study.

\begin{figure}
 \begin{center}
   \includegraphics[width=8cm]{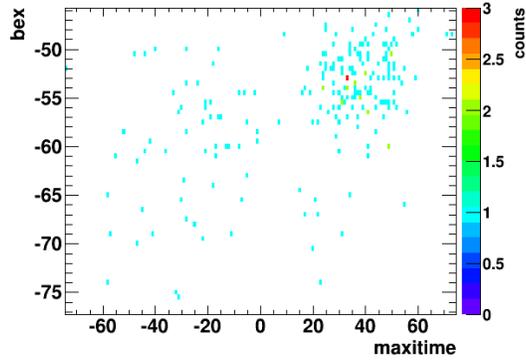}	
 \end{center}
\caption{Event distribution in  the detector coordinate
extracted for the PSF-fit for V5586 Sgr on a scan at MJD 55340 
by a camera (ID = 0) in the $2-4$ keV band.
Horizontal and vertical axes are the relative time(s)
from 2010-05-24 15:34:14 (UT) and bex (mm), respectively.
 The color scale shows the number of events in a bin.}
\label{fig:selevt}
\end{figure}

We select events from a rectangular region
in the detector coordinate,  where the width of time (scan)
direction is determined  to be three times the width
of  the base side of the triangular shape.
That of the anode direction is set 
 to be 16 sigma of the Gaussian shape.
 As a result, these widths vary in every scan.
We then add a constraint for these widths
so that the rectangular region cannot
exceed a circular sky region
with a radius of 8 deg, which is the radius of
the event extraction at the beginning of the analysis
(Section \ref{sec: obs ana}).
Figure \ref{fig:selevt} shows an example of
the event distribution in  the detector coordinate to
be used for the PSF-fit. It is for V5586 Sgr on a scan at 
the modified Julian day (MJD) 55340 
 with a camera (ID = 0) in the $2-4$ keV band.

We minimize the following cost function, c-stat
\citep{Cash 1979}, to calculate the fluxes of 
a target and nearby sources,
\begin{eqnarray}
  c & = & \sum_{k \in A} c_k = -2 \sum_{k \in A} \ln L_k \nonumber \\
  & = & -2 \sum_{k \in A} \left[ \sum_{i \in E_k}
    \ln F_{\rm model}(t_i, b_i; k) \right. \nonumber \\
    && - \left. \int_{\rm M_k} F_{\rm model}(t, b; k) dt\,db \right],
\end{eqnarray}
 where $A$ is a set of the operating cameras and $E_k$ is 
a set of the events detected in a camera ($k$).
$M_k$ is an integration area for a camera ($k$),
where the time duration of HV-off and
the center of  the camera in  the anode direction ($|b| < 4$~mm) are masked out.
The region of the anode edges ($|b| > 130$~mm)  is also masked out.
To speed up the calculation,
we use unbinned likelihood instead of binned likelihood.
\texttt{Minuit2}
\footnote{http://seal.web.cern.ch/seal/snapshot/work-packages/mathlibs/minuit/}
minimizer is used for  minimization of the cost function.
To calculate the error ranges, we calculate
the likelihood interval \citep{Cowan 1998}, 
using \texttt{minos} in \texttt{Minuit2}.

When a target  is located in a crowded region like the Galactic plane,
there are many various nearby sources;
some sources are always bright enough to be
detected by MAXI/GSC, while some are  usually faint and
have a potential to become bright above the
threshold of MAXI/GSC sensitivity.
For almost  any region in the sky,  MAXI/GSC
 is the only X-ray detector that monitors these sources; then
it is impossible to determine a priori
which sources are detectable and which are not.
 However, we must select which source is necessary
to be included in the model for the PSF-fit.
 We approach this problem as follows.

\begin{figure}
 \begin{center}
   \includegraphics[width=8cm]{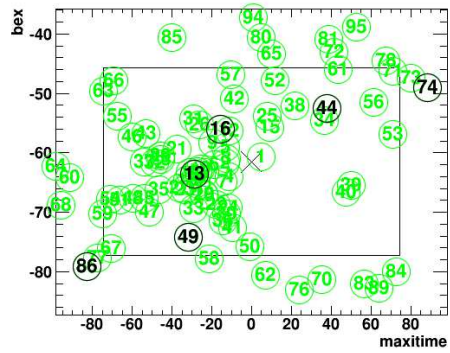}	
 \end{center}
\caption{The positions of the nearby source candidates
and the selected sources are shown in numerical labels
with green and black colors, respectively.
The cross label at the center
is the position of the target, V5586 Sgr. The scan time and camera
are the same as in Figure \ref{fig:selevt}.
The area used for the PSF-fit is  displayed as a rectangular region,
which is the same area as that  in Figure \ref{fig:selevt}.
}\label{fig:objs}
\end{figure}

We pick up candidates for the
nearby sources to be included in the PSF-fit
from the  catalog 
used in the Nova search system \citep{Negoro+2015}.
The area for picking up these candidates
are the rectangular region in the detector coordinate
as described above and the outer marginal region
with half width of  the base side of the triangular shape
for the time direction 
and three sigma of the Gaussian shape for the anode direction.
This marginal region for selecting the nearby sources
is necessary to account for the contamination of
PSFs of sources out of the fitting region,
within which the events are extracted
(Figure \ref{fig:selevt}).
Figure \ref{fig:objs} demonstrates the positions of
the nearby candidate sources listed in the catalog
in the example case, as well as
the region for the event extraction (fitting) 
and outer marginal regions. 

The nearby sources included in the PSF-fit
(black labels in Figure \ref{fig:objs})
are selected  with the measure of the Bayesian information criterion
(BIC; \cite{Schwarz 1978}) from the candidates
(green labels in Figure \ref{fig:objs}) as follows.
First, we fit the event distribution  with
 the simplest model  that includes only the background,
and calculate the BIC.
Next, we include one nearby source among
all the nearby candidate sources into the model function,
and fit the events  with this model and calculate the BIC.
We then identify the best nearby source with the minimum BIC
as the first selected nearby source.
If this minimum BIC is larger than that of the previous BIC
obtained by the fit  that includes only the background,
which means that no nearby source is
necessary to improve the fit, then
the model selection for the nearby sources  is completed.
Otherwise, we include this source in the final model,
then proceed to the next step.
In the next step, we include another nearby source
among the rest of the nearby sources, and repeat the above procedure.
We stop this procedure when the BIC reaches the minimum.
Finally, we add the target source in the final model.
Figure \ref{fig:objs}  demonstrates the positions of nearby sources
selected  with this model selection, and these 
are used for the fitting process.

\begin{figure}
 \begin{center}
   \includegraphics[width=8cm]{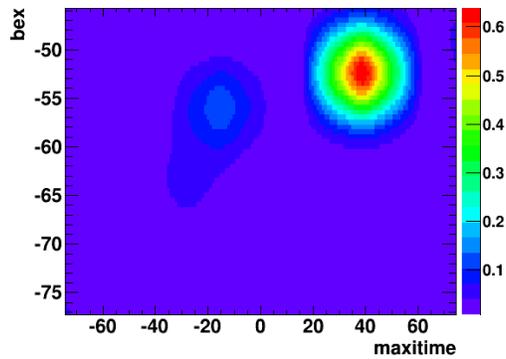}	
 \end{center}
\caption{The best-fit function obtained  with the PSF-fit
for the same region as in Figure \ref{fig:selevt}.
The value of the best-fit function  are shown in  the unit
of counts s$^{-1}$ mm$^{-1}$ (color bar).
}\label{fig:func_best}
\end{figure}

\begin{figure}
 \begin{center}
   \includegraphics[width=5cm, angle=-90]{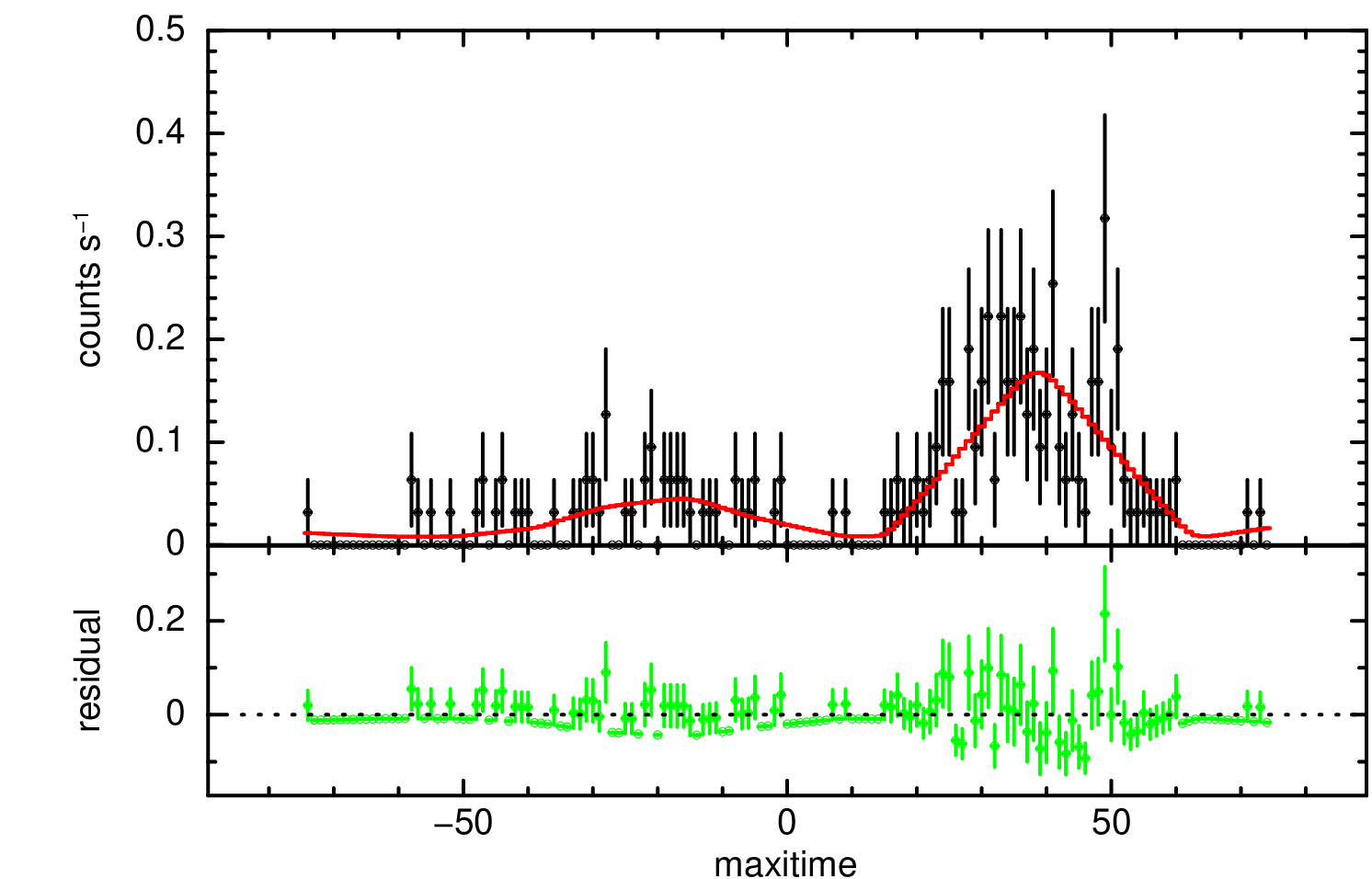}	
 \end{center}
\caption{(Top panel:) The projection of the best-fit function
 in Figure \ref{fig:func_best} onto the time direction
(red histogram). The same projection of the event distribution 
of Figure \ref{fig:selevt} is  given in crosses.
The vertical axis is the event rate in  the unit of counts s$^{-1}$.
(Bottom panel:) The residual of  the best\textcolor{red}{-}fit.
}\label{fig:projx}
\end{figure}

\begin{figure}
 \begin{center}
   \includegraphics[width=5cm, angle=-90]{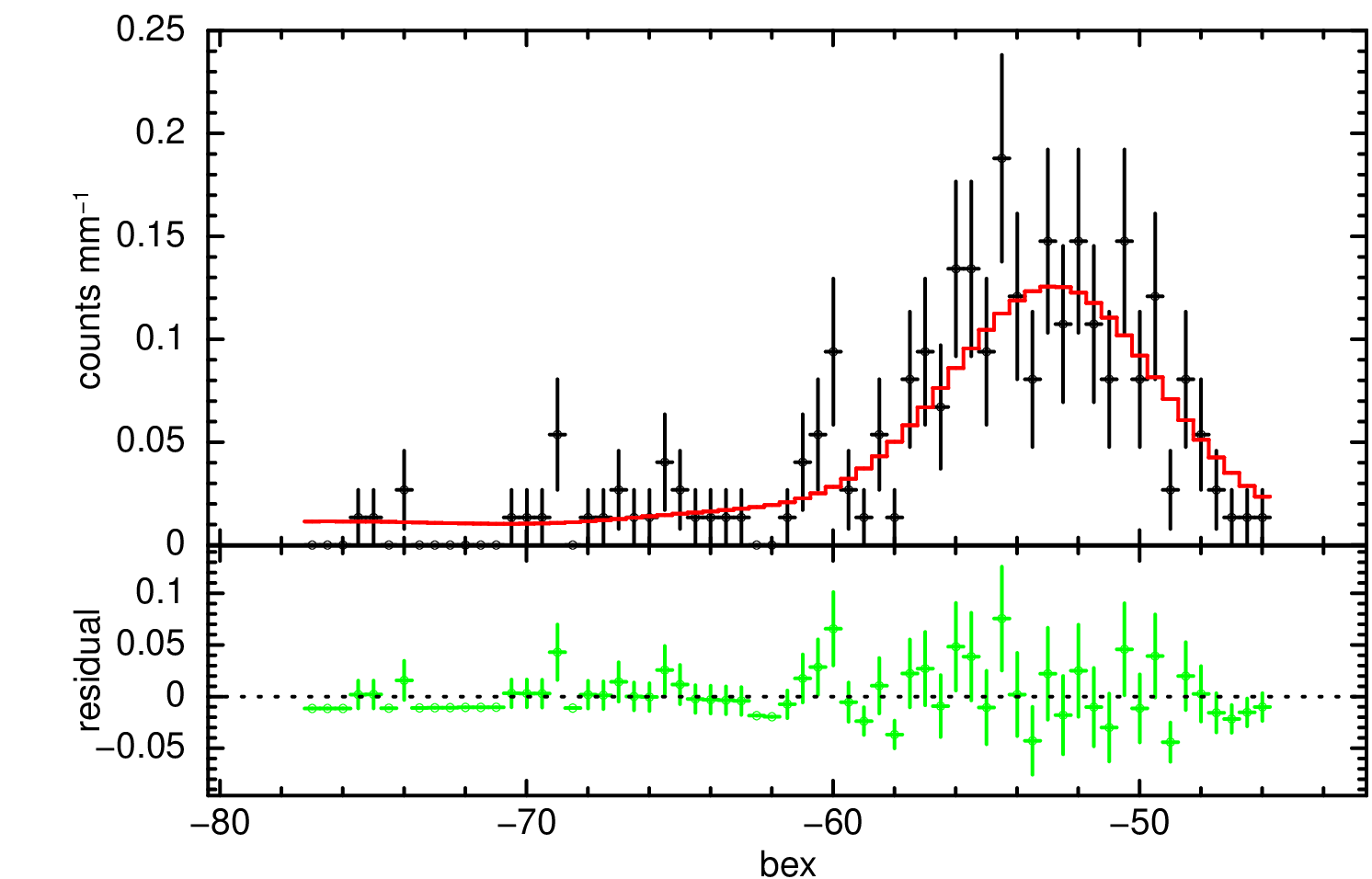}	
 \end{center}
\caption{(Top panel:) The projection of the best-fit function
 in Figure \ref{fig:func_best} onto the anode direction
(red histogram). The same projection of the event distribution
of Figure \ref{fig:selevt} is  given in crosses.
The vertical axis is the event rate in  the unit of counts mm$^{-1}$.
(Bottom panel:) The residual of  the best\textcolor{red}{-}fit.
}\label{fig:projy}
\end{figure}

Figures \ref{fig:func_best}, \ref{fig:projx} and \ref{fig:projy} show
an example of the best-fit model obtained
 with the PSF-fit tool;
 they show the model function,  and
the projection onto the time and anode directions,
 respectively.

\begin{figure}
 \begin{center}
   \includegraphics[width=6cm, angle=-90]{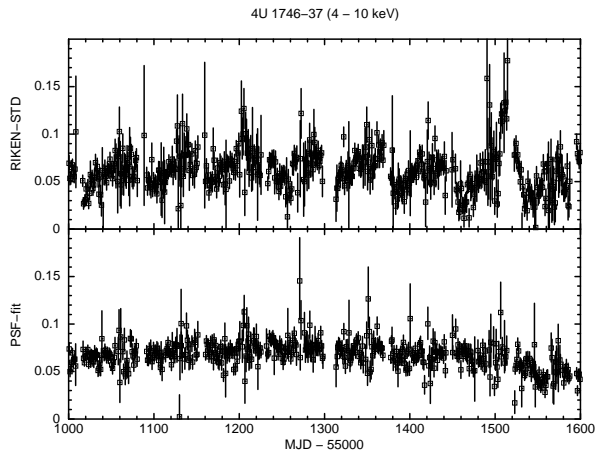} 	
 \end{center}
\caption{The light curves of 4U 1746-37 in the $4-10$ keV band in one-day bins
in  the unit of counts s$^{-1}$ cm$^{-2}$. Horizontal axis is in days (MJD - 55000).
The top panel shows the light curve in the data archive at the RIKEN web site,
 whereas the bottom panel is that obtained  with the PSF-fit.
}\label{fig:J1750}
\end{figure}

To demonstrate the efficacy of the PSF-fit, 
we present Figure \ref{fig:J1750}.
It shows a light curve of a moderately bright
and stable source 4U 1746$-$37.
Due to a periodic contamination of the tail of
 the PSF of a nearby bright source, the light curve
 available  on the website of RIKEN is disturbed 
 with a periodic artificial modulation 
caused by the precession of the ISS (top panel).
 With the PSF-fit, this modulation effect is clearly removed (bottom panel).

\end{document}